\documentclass[lettersize,journal]{IEEEtran}
\usepackage{amsmath,amsfonts}
\usepackage{algorithmic}
\usepackage{array}
\usepackage[caption=false,font=normalsize,labelfont=sf,textfont=sf]{subfig}
\usepackage{textcomp}
\usepackage{stfloats}
\usepackage{url}
\usepackage{verbatim}
\usepackage{graphicx}
\usepackage{orcidlink}
\def\BibTeX{{\rm B\kern-.05em{\sc i\kern-.025em b}\kern-.08em
    T\kern-.1667em\lower.7ex\hbox{E}\kern-.125emX}}
\usepackage{balance}

\begin{document}

\title{Assessing the operating characteristics of an ion-milled phonon-mediated quantum parity detector}

\author{
    Brandon J. Sandoval\,\orcidlink{0009-0008-7671-9472},
    Andrew D. Beyer\,\orcidlink{0000-0002-1045-1652},
    Pierre M. Echternach\,\orcidlink{0000-0001-9643-5110},
    Sunil R. Golwala\,\orcidlink{0000-0002-1098-7174},
    William D. Ho\,\orcidlink{0009-0002-8675-9212},
    Lanqing Yuan\,\orcidlink{0000-0003-0024-8017},
    and Karthik Ramanathan\,\orcidlink{0000-0003-4215-7834}
    
    \thanks{The work was supported in part by National Science Foundation PHY Grant 2209581 and DARPA QuSeN Grant HR00112520037. Work was conducted at the California Institute of Technology, NASA’s Jet Propulsion Laboratory, and Washington University in St. Louis (Corresponding author: Karthik Ramanathan.)}
    \thanks{W. D. Ho, L. Yuan, and K. Ramanathan are with the Department of Physics, Washington University in St. Louis, St. Louis, MO, USA.}
    \thanks{B. Sandoval and S. Golwala are with the Division of Physics, Mathematics, and Astronomy, California Institute of Technology, Pasadena, CA, USA.}
    \thanks{P. M. Echternach and A. D. Beyer are with NASA Jet Propulsion Laboratory, Pasadena, CA, USA}
}

%\markboth{IEEE Transactions on Applied Superconductivity,~Vol.~X, No.~Y, January~2026}%
%{B. Sandoval, \MakeLowercase{\textit{(et al.)}}}

\maketitle

\begin{abstract}
Phonon sensitive superconducting qubits promise to provide sub-eV energy deposit thresholds, useful for future rare-event experiments looking for interactions from dark matter and neutrinos. We detail here engineering results from a Quantum Parity Detector (QPDs), one of a class of phonon sensitive qubits, and, as a first measurement, show that this device has a quiescent quasiparticle density of $1.8 \pm 0.8 \mu \mathrm{m}^{-3}$, in line with expectation. We also outline an argon ion-mill process for multi-step Josephson Junction fabrication, expanding the sparse literature on this topic, which proves useful in avoiding secondary parasitic junctions.
\end{abstract}

\begin{IEEEkeywords}
Calorimetry, dark matter, neutrinos, phonons, quantum parity detector, qubit, superconducting, RF
\end{IEEEkeywords}

\section{Introduction} \label{sec:intro}
\IEEEPARstart{R}{are}-event search experiments look for signatures from exceedingly infrequent particle interactions, such as those created by dark matter (DM) or neutrinos ricocheting off nuclei \cite{hooper2008strategies, freedman1974coherent}. A technically challenging, but well motivated, area for exploration is when the deposited energy from these interactions is at the eV-scale and lower \cite{kolb2018basic}. Below deposited energies of $\mathcal{O}$(1~eV) for electron recoils and $\mathcal{O}$(10~eV) for nuclear recoils, the transferred energy is insufficient to ionize electrons or unbind nuclei for most common crystalline targets, hence detector modalities relying on charge production and/or scintillation are no longer reliable \cite{golwala2022novel}. However, collective excitations such as phonons (lattice vibrations in a crystal) can be robustly excited by such interactions and have complementary energy scales \cite{trickle2020multi}. As such, development of sub-eV threshold phonon-mediated detectors, though challenging, would allow for the probing of new physics in previously unexplored parameter space.

\begin{figure}[!t]
\centering
\includegraphics[width=\linewidth]{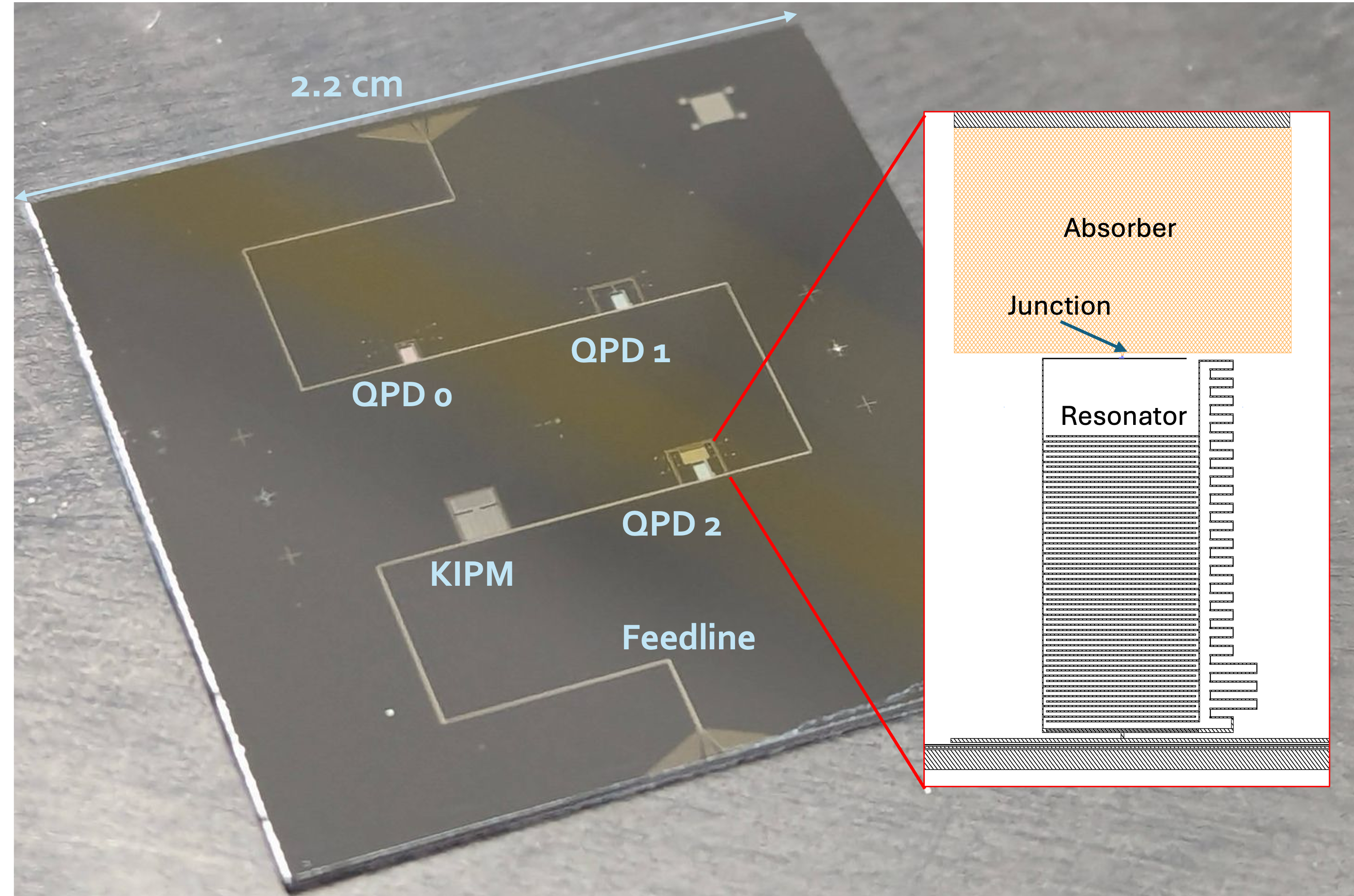}
\caption{Labeled optical image of 1~mm thick silicon chip with 4 phonon sensitive devices (3 QPDs and 1 KIPM. \textit{Inset: } Schematic of a single QPD, see Fig. \ref{fig:milledjunction} for a close-up of the Josephson junction.}
\label{fig:chip}
\end{figure}

In parallel, the last decades have seen an explosion in
quantum computing — the field of using quantum phenomena like entanglement and interference to solve certain hard computational problems faster than classical approaches. The central element here is a qubit, a two-level quantum mechanical system, with physical implementations encoding
information in structures as diverse as photons, cold atoms, and superconducting circuits, among others. An example of the latter is a Cooper-pair Box (CPB) \cite{wallraff2004strong}, wherein the system's quantum states correspond to specific numbers of excess charges in a localized region. These qubits saw great interest in the early 2000s as a potential quantum computing platform but fell out of favor because they were unstable to environmentally induced charge noise (i.e. quasiparticle poisoning) \cite{martinis2009energy}. 

However, this sensitivity to charge noise, which manifests as a shift in the qubit's charge parity, can be exploited to build a detector. Quantum Capacitance Detectors (QCDs) are photon counters based on CPBs and have shown the ability to count single 1.5 THz photons \cite{shaw2009quantum, echternach2018single}. Ideas to extend qubits to phonon detection have recently been proposed in the literature \cite{ramanathan2024quantum,fink2024superconducting,linehan2025estimating}, including by the authors, and we specifically investigate Quantum Parity Detectors (QPDs) which are an evolution of QCDs engineered for phonon sensitivity. 

In this work, we briefly characterize the performance of one of the first QPD chips to be fabricated (seen in Fig. \ref{fig:chip}), consisting of 3 QPD sensors and a Kinetic Inductance Phonon Mediated (KIPM) detector \cite{temples2024performance}, an alternative phonon-mediated device derived from Kinetic Inductance Detectors \cite{day2003broadband}. Section \ref{sec:operation} briefly lays out the relevant design and theory of operation of our QPDs. Sec. \ref{sec:mill} details unique fabrication techniques we have used to make these devices more useful for particle detection. Sec. \ref{sec:performance} outlines realized performance. Sec. \ref{sec:qpdensity} presents a key measurement, the quiescent (i.e. background) quasiparticle density within a QPD absorber. Finally we comment briefly on KIPM performance in Sec. \ref{sec:KIPM}, similarly extracting a quiescent quasiparticle density and showing it as significantly elevated compared to a QPD.

\section{Device Design and Theory of Operation} \label{sec:operation}

\begin{figure}[!h]
\centering
\includegraphics[width=\linewidth]{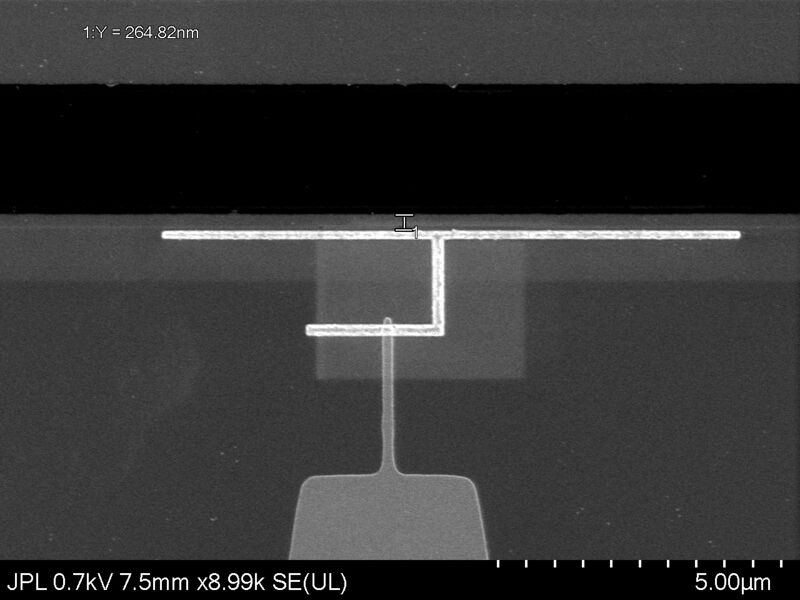}
\caption{SEM image of ion-milled junction. Bright (J-shaped) structure is the secondarily deposited island, overlaying the darker (grey) first deposited absorber and lead. The intersection forms the Josephson junction. Above the island lies the resonator (black rectangle).}
\label{fig:milledjunction}
\end{figure}

A QPD is composed of 5 elements: a substrate, absorber, Josephson junction, island, and readout resonator. The substrate is the macroscopic crystalline target for a particle interaction. Said interaction produces a burst of $\mathcal{O}$(THz) phonons that eventually propagate quasi-ballistically through the crystal. The phonons enter thin $\mathcal{O}$(100)~nm superconducting absorber films patterned on the surface of the substrate and break Cooper-pairs  (pair-breaking energy $2\Delta$) within this film. These resultant quasiparticles can tunnel, via narrow leads, through the central Josephson junction (a $\mathcal{O}$(100)~nm square Superconductor-Insulator-Superconductor sandwich) onto a small $\mathcal{O}$($\mu$m$^3$) superconducting island. The entire structure is capacitively coupled to a superconducting LC resonator. A labeled micrograph of a QPD can be found in Fig. \ref{fig:chip} and inset. The junction \& islands are more clearly seen in Fig. \ref{fig:milledjunction}. The three QPDs have identical junctions but with absorber sizes spanning 1, 100 and 1000 $\rm{\mu m^3}$ (QPD0, 1, 2 respectively), in an effort to quantify the volume dependency of their responsivity. 

The repeated tunneling of quasiparticles modifies the energy-level structure of the qubit \textemdash\ shifting the ground state energy of the qubit between parity states $P=\pm1$, referred to as \textit{even} and \textit{odd}. This change of energy with respect to charge can be interpreted as a changing capacitance, engendering a measurable $\mathcal{O}$(MHz) binary shift in the LC resonator's $\mathcal{O}$(GHz) frequency. 

With the exception of the substrate/absorber coupling, the design of a QPD is near-identical to a QCD and details of the latter can be found in Refs. \cite{echternach2018single, echternach2021large}, including specifics about parity switching, film thicknesses, feedline design, and the equations describing the qubit energy levels. In general though, there are 3 qubit specific parameters that one can tune: $E_c$, a charging energy related to moving a Cooper-pair onto the island; $E_J$, the Josephson energy of the junction related to the size and thickness of the junction; and $n_g$ (or equivalently $V_g$), an externally modulated voltage induced charge on the island.

The key operating principle for these QCDs/QPDs to work as a detector is that the tunneling rate from the absorber to the island is linearly proportional to the density of quasiparticles in the absorber, $\Gamma_{\rm in}=Kn_{\rm qp}$. The tunnel out rate $\Gamma_{\rm out}$, from island back to absorber, is effectively constant \cite{lutchyn2006kinetics}. For our designs, under the assumption that the absorber is not diffusion limited and further that the gap energy is symmetric across the junction, the tunneling constant can be analytically written \cite{shaw2009quantum} as,
\begin{align} \label{eq:K}
    K &= \frac{G_N}{e^2}\frac{e^{\Delta/k_BT}}{N_0\sqrt{2\pi\Delta k_B T}}\int_{\Delta}^{\infty} dE h(E) e^{-E/k_B T} \\
    h(E) &= \frac{E(E+\delta E)-\Delta^2}{\sqrt{((E+\delta E)^2-\Delta^2)(E^2-\Delta^2)}} \nonumber
\end{align}
where $G_N$ is the junction conductance, $T$ is the operating temperature, $N_0$ is the single spin density of states at the Fermi level, and $\delta E$ is the difference in energy between the parity states. Crucially then, having computed $K$, one can infer the background quasiparticle density within the absorber by measuring the steady-state $\Gamma_{\rm in}$. 

\section{Fabrication and Ion-milled Junctions} \label{sec:mill}

Standard junction patterning techniques, such as by shadow evaporation, inherently produce parasitic secondary junctions due to the overlap created by consecutive metal, oxide, metal deposition (without breaking vacuum) within the high-vacuum deposition tool. These unwanted junctions can inadvertently act to blockade quasiparticle transport within the absorber, thereby reducing QPD sensitivity. 

\begin{figure}[!h]
\centering
\includegraphics[width=\linewidth]{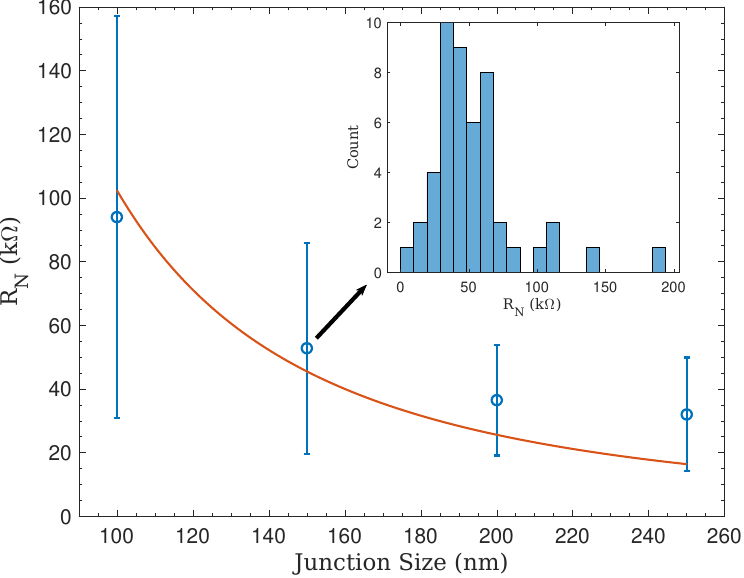}
\caption{Normal state resistance as measured at room temperature vs square junction side length. The overlaid (red) line is a fit to a 1/(junction area) model; The inset histogram shows the distribution of measured resistances for the 150 nm side length test array.}
\label{fig:resistanceplot}
\end{figure}

To circumvent this issue, we adopted a non-standard junction patterning process inspired by Refs. \cite{wu2017overlap,grunhaupt2017argon}, where junction patterning is done in multiple steps. We first e-beam evaporated an aluminum ($40$~nm) layer on a virgin Si wafer (double-side polished, $>10\,\rm{k}\Omega\cdot\rm{cm}$, 1~mm thick), initially cleaned with a Buffered Oxide Etch. Next, we stenciled a PMMA/DUV resist mask and used a chlorine based RIE to define the absorber and bottom electrode. We then patterned and deposited the Nb readout resonators (ranging from 2.8-3.2 GHz in resonant frequency) and ground plane via lift-off. The intermediate air exposures promote uncontrolled oxide growth on the bottom electrode, anathema to properly constructing the central junction. To remove this, we defined the top electrode and island using e-beam lithography and introduced the wafer back into the Al deposition chamber where a calibrated Ar ion mill (operated at 30~mA, 300~V) stripped this oxide. We estimate up to $\sim$15~nm of bottom electrode material was also removed. We then flowed O$_2$ (15 mins at 100 mT) to create a controlled AlOx barrier, before final Al deposition and liftoff steps.

We constructed a series of 50 test junctions of side lengths ranging from $100~\rm{nm}$ to $250~\rm{nm}$, while sweeping parameters such as the RF mill power and oxidation times and pressures. We measured their normal state resistivity, $R_N$ ($\equiv1/G_N$), which informs the expected QPD tunneling rate as per Eq. \ref{eq:K}. Figure \ref{fig:resistanceplot} shows both the average and spread of resistances for our current parameter choices, with the inset showing the distribution of resistances for a 150~nm square junction. We are able to achieve $R_N$ of approx. $\mathcal{O}(10)~\rm{k}\Omega$, as needed for charge qubit operation. However, the significant spread in junction resistances, on the order of 50\%, suggest further process requirement is required to hit specific target $R_N$ with low variability. 

\section{Device Operation} \label{sec:performance}

We packaged the QPD chip in an SMA connected gold plated copper box, similar to the ones used in Ref. \cite{temples2024performance}. The holder was placed within a Stycast coated copper can, mounted on the base stage of a Bluefors LD250 dilution refrigerator, and subsequently cooled down to 9~mK. Input tones, at -95~dBm at the device after about 70~dB of cold attenuation, were provided by a room temperature signal generator, and tied via a bias-tee to a function generator provided $V_g$. Output signals from the QPD were amplified by a 4~K stage HEMT amplifier before further warm stage amplification and eventual DAQ capture at a 100~kHz sampling rate via a homodyne mixer scheme. The operational environment, including RF wiring, shielding, and readout electronics were similar to prior QCD runs and details, again, can be found across Refs. \cite{echternach2018single, echternach2021large} .

\begin{table}[!ht]
\caption{Extracted resonator parameters for the 4 on-chip sensors, descriptions in text. Statistical uncertainties on quality factors are at the \% level.}
\centering
\resizebox{\columnwidth}{!}{\tiny%
\begin{tabular}{|l|l|l|l|}
\hline
\textbf{Sensor} & \textbf{$\rm{f_r}$ (GHz)} & \textbf{$\rm{Q_i}$} & \textbf{$\rm{Q_c}$} \\ \hline
KIPM & 4.040 & 277000 & 12500 \\ \hline
QPD0 & 3.032 & 34100  & 6720  \\ \hline
QPD1 & 2.999 & 24300 & 2020 \\ \hline
QPD2 & 2.857 & 2920 & 2600 \\ \hline
%KIPM & 4.040 & 11957 & 277213 & 12496 \\ \hline
%QPD0 & 3.032 & 5612 & 34072  & 6719  \\ \hline
%QPD1 & 2.999 & 1868 & 24290 & 2024 \\ \hline
%QPD2 & 2.857 & 1373 & 2915 & 2597 \\ \hline
\end{tabular}%
}
\label{tab:resparams}
\end{table}

The QPDs were first characterized by a VNA scan, and resonator parameters were extracted using the tool outlined in Ref. \cite{probst2015efficient} (see Table \ref{tab:resparams}; $1/\rm{Q_r}=1/\rm{Q_i}+1/\rm{Q_c}$, with overall, internal, and coupling quality factors respectively, along with high-power resonant frequency $f_r$) and accounting for potential impedance mismatches \cite{khalil2012analysis}. The coupling quality factors had a design target of $\approx10000$.

Next, to check if the devices were alive, we conducted a slow $V_g$ DC bias sweep (up to 3~mV at the device, based on circuit calculations) while monitoring for increases in the variance of the $S_{21}$ transmission quadratures. Of the three QPDs, only the highest frequency, smallest absorber volume device, QPD0, was determined to be responsive. The other two sensors did not respond to any DC bias or increased readout power, suggesting either the junctions were non-functional or had a poor resonator coupling.  

Minute long timestreams of QPD0 were recorded under two operating schemes: with fixed bias at the maximal parity separation point ($V_g\approx1\,\rm{mV}$ such that $n_g$=0.5) and with a fast 5~kHz sawtooth sweep of $n_g$ from 0 to 1. This latter operating mode nominally allows for multiple sensors to be read out using the same feedline, as no two sensors will have the same optimal bias point \cite{echternach2021large}.

\begin{figure}[!ht]
\centering
\includegraphics[width=\linewidth]{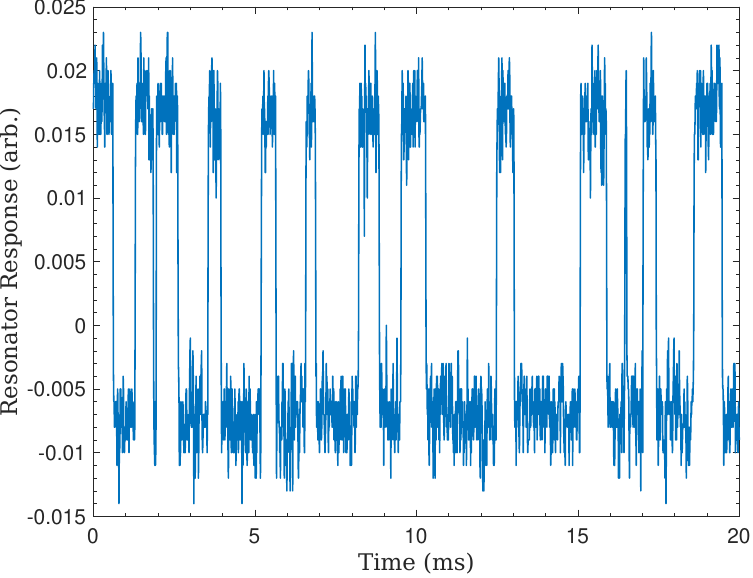}
\caption{Example of a QPD0 tunneling trace, operated with a constant DC bias such that $n_g\approx0.5$, showing the telegraph noise behavior arising from switching between even (low) and odd (high) parity states.}
\label{fig:tunnelingrate}
\end{figure}

A brief snapshot of the transmission trace of a single readout quadrature of QPD0, under fixed bias, can be seen in Fig. \ref{fig:tunnelingrate}. This plot clearly resolves the binary switching of the QPD between even and odd ground states. Relatedly, Fig. \ref{fig:multitrace} shows the result of the swept bias, referred to as the capacitance trace, demonstrating the expected sinusoidal behavior \cite{echternach2018single}. These results provide confirmation that the QPD operational concept is sound and continued work can focus on optimizing other parts of the sensor chain, including the phonon collection efficiency of the absorber.

\section{Extracting Quiescent Quasiparticle Density} \label{sec:qpdensity}

\begin{figure}[!t]
\centering
\includegraphics[width=\linewidth]{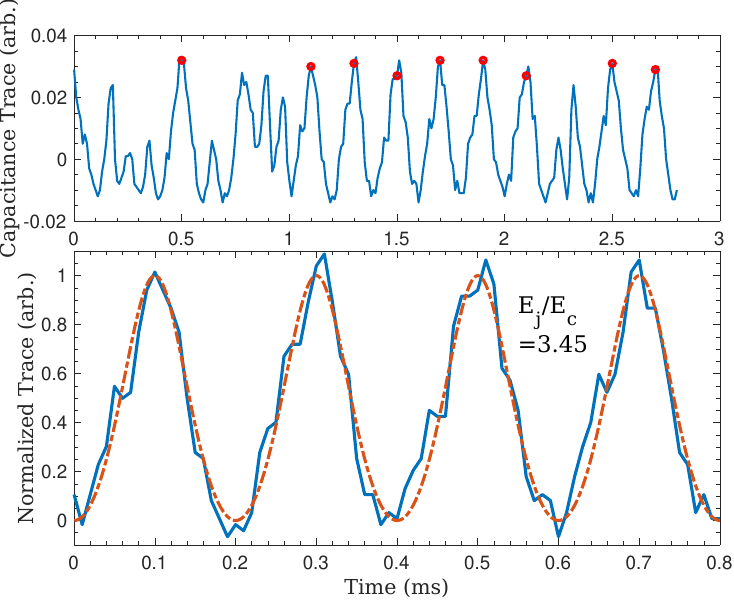}
\caption{\textit{Top:} A few ms portion of a quantum capacitance trace, constructed by sweeping the DC bias through a full period of $n_g$ charge and measuring the resonator $S_{21}$ quadrature responses. A characteristic sinusoidal pattern is observed, interrupted by bursts of excess quasiparticles (e.g. arising from dark count events) spoiling the trace. Select peaks (partial sample highlighted in red) are pulled from the dataset to construct a template, and this is fit by a one parameter energy ratio $E_j/E_c$ model that determines the shape of the qubit energy levels. \textit{Bottom: } A best-fit energy ratio of $\sim$3.45 is extracted from this template and is shown overlaid on a select few peaks.}
\label{fig:multitrace}
\end{figure}

A rare-event search capable QPD will require that the quiescent (background) quasiparticle density within the phonon absorber be $\ll1$~$\rm{\mu m^{-3}}$ \cite{ramanathan2024quantum}. Al qubit quasiparticle densities in the literature are reported to be at the $0.1$\textendash$1\,\rm{\mu m^{-3}}$ level \cite{connolly2024coexistence,saira2012vanishing,riste2013millisecond}. To determine how far current QPDs are from achieving this metric, we extract the quasiparticle density using Eq. \ref{eq:K} in conjunction with the tunneling traces presented in Sec. \ref{sec:performance}. However, computing Eq. \ref{eq:K} is not straightforward as it requires accurate estimation of $E_J$ and $E_C$.  

In principle, as these QPDs are charge qubits, one can attempt to find the specific $|0\rangle \to |1 \rangle$ transition energies as a function of $n_g$ and reconstruct the energies. However, the expected qubit transition frequency is on the order of 20~GHz, which is not easily accessible in our setup. In addition, as introduced in Sec. \ref{sec:intro}, charge qubits are very unstable (with $\mathcal{O}$(ns) coherence times), making the whole exercise fraught with complications. 

An alternative method is to directly measure $E_J$ or $E_C$. Yet here again, as the Josephson Junction leads are much smaller than any probe station probes, extracting $E_J$ after fabrication via a resistance measurement and using the Ambegaokar-Baratoff relations \cite{ambegaokar1984quantum} is near impossible. Furthermore, the decoupling of the junction from direct charge injection, unlike a Single Electron Transistor (SET) design, means that $E_C$ cannot be measured through IV curve measurements. In principle $E_C$ can be estimated through simulation, but is prone to significant systematics arising from parasitic capacitances or imperfect couplings. 

With these complexities in mind we instead use the shape of the swept bias capacitance trace, which tracks the curvature of the ground state energy level, to determine $E_J/E_C$ (see Ref. \cite{shaw2009quantum} for explicit relations). Using a 5~s portion of a capacitance trace, we algorithmically identify clean peaks in the trace, as seen by the circles in Fig. \ref{fig:multitrace} Top. Deviations from a periodic trace arise from rapid tunneling after energy deposits create a burst of quasiparticles (e.g. dark count events) \cite{echternach2018single}. We fit an averaged ($N\sim13100$) single period curve template with an analytically determined trace, resulting in a best-fit value of $E_J/E_C=3.45\pm0.1$. This is significantly larger than prior QCD sensors ($E_J/E_C<1$) and is plausibly due to a much smaller than expected $R_N$ from our ion-milled fabrication process. This extracted ratio, in conjunction with an estimation of $E_J$ from $R_N$ through Fig. \ref{fig:resistanceplot}, allows us to compute a distribution for $K_{\rm data}$ from Eq. \ref{eq:K}.

\begin{figure}[!ht]
\centering
\includegraphics[width=\linewidth]{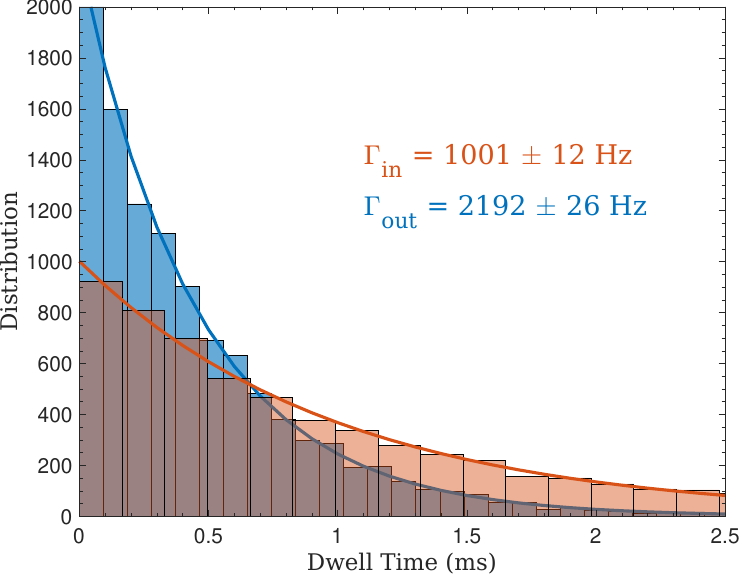}
\caption{Dwell time distributions computed from tunneling traces, with overlaid fits and resultant tunneling rates.}
\label{fig:dwellfits}
\end{figure}

Next, we estimate $\Gamma_{\rm in}$ from the fixed bias data. QPD operation theory, and prior experience with QCDs, suggests that the dwell times of the tunneling trace will be asymmetric, with longer dwell times in the even (0 or an even number of quasiparticles on the island) state \cite{shaw2009quantum}. We extract the tunneling rate from fitting simple one parameter exponentials to both the high and low dwell time distribution of a minutes long tunneling trace, as seen in Fig. \ref{fig:dwellfits}, resulting in an estimate of $\Gamma_{\rm in}^{\rm data }\approx 1$~kHz. 

\begin{figure}[!h]
\centering
\includegraphics[width=\linewidth]{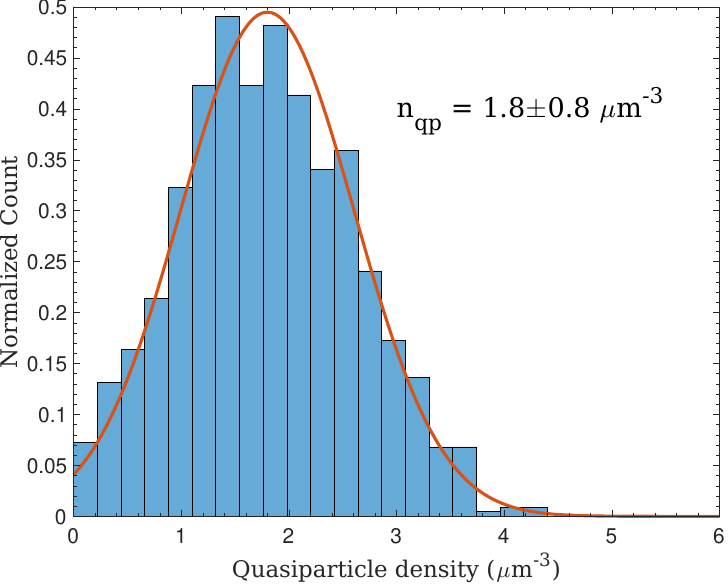}
\caption{Extracted quasiparticle density in the QPD0 absorber, determined via Monte-Carlo draws, with an overlaid normal distribution fit (red line). We report a quiescent quasiparticle density of $n_{\rm qp}\sim1.8\pm0.8\mu m^{-3}$.}
\label{fig:Kmontecarlo}
\end{figure}

Putting together the different threads, we compute $n_{\rm qp}$ as $\Gamma_{\rm in}^{\rm data}/K_{\rm data}$. We use a Monte-Carlo approach, flowing through statistical errors from our prior estimations, and arrive at the distribution shown in Fig. \ref{fig:Kmontecarlo}. We extract a best fit of $n_{\rm qp}=1.8\pm0.8$~$\rm{\mu m^{-3}}$, which is reasonably in line with aforementioned literature sources \cite{connolly2024coexistence,saira2012vanishing,riste2013millisecond}. This further validates our approach to constructing QPD sensors, though future work will require a dedicated campaign to investigate and mitigate potential sources (e.g. parasitic IR radiation \cite{malevannaya2025}) of the quiescent population. 

\section{Co-calibration KIPM sensor} \label{sec:KIPM}

Finally, our QPD chips to date have also included a secondary type of sensor, a KIPM. Briefly, a KIPM exploits kinetic inductance \textemdash\ an inductance related to the inertia of Cooper-pairs to applied AC fields \textemdash\ to measure shifts in quasiparticle density. KIPM sensors have shown great promise for rare-event detection, with projected eV-scale thresholds \cite{wen2022performance,ramanathan2022identifying}, and have been well calibrated \cite{temples2024performance, cardani2021final}, which should allow for cross-calibration with the less well-studied QPD architecture. 

\begin{figure}[!h]
\centering
\includegraphics[width=\linewidth]{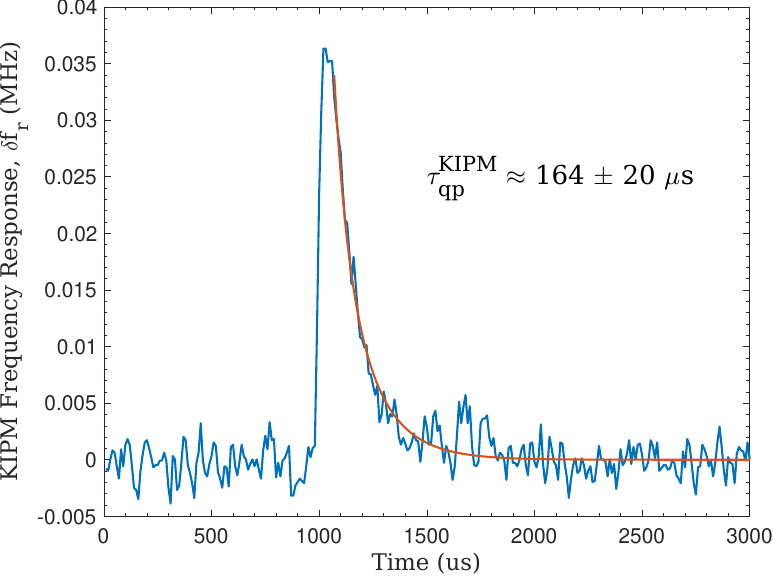}
\caption{Sample pulse as measured by an on-chip KIPM similar to the one seen in Fig. \ref{fig:chip}. The change in quasiparticle density, likely induced a cosmic-ray, shifts the resonant frequency of the device, which can be traced as a pulse. The solid overlay (red) denotes a physically motivated exponential fit to the fall of the pulse, extracting a quasiparticle lifetime of approx. 165 $\mu$s. Note, this specific pulse comes from a secondary KIPM in an identical chip to the one primarily discussed in text. This alternate chip did not have stable QPDs, potentially due to fabrication issues, but did present with larger KIPM responsivity.}
\label{fig:kidpulse}
\end{figure}

The Al KIPMs on our chips were also defined through the first 40~nm Al deposition step oultined in Sec. \ref{sec:mill} and were designed with a $\sim24000\,\rm{\mu m^{3}}$ inductor volume. They were read out with feedline power $P_{\rm read}\approx-75$~dBm. We estimate the quasiparticle density in the KIPM volume by studying cosmic ray events, an example of which is shown in Fig. \ref{fig:kidpulse}.

Following the prescription in Ref. \cite{de2021strong}, we compute

\begin{align} \label{eq:kiddensity}
    n_{\rm qp}^{\rm KIPM} &= \frac{\tau_0 N_0 (k_B T_c)^3}{2\Delta^2 \tau_{\rm qp}^{\rm KIPM}} \approx 710\,\rm{\mu m^{-3}},
\end{align}
which contains the quasiparticle lifetime $\tau_{\rm qp}^{\rm KIPM}$, and $\tau_0$ is a material-dependent characteristic time for the electron-phonon coupling. We use $\tau_{\rm qp}^{\rm KIPM}\approx 165\,\rm{\mu s}$, as extracted from fitting to pulse fall-times (see Fig. \ref{fig:kidpulse} again) using a model that accounts for quasiparticle generation-recombination \cite{de2014fluctuations}. Even allowing for secondary effects like phonon escape and mis-estimation of parameters, this computed quasiparticle density is at least a 100$\times$ larger than the qubit derived one. This elevated density is arguably sourced by readout power, through complex thermal and multi-photon processes \cite{de2014fluctuations,chowdhury2025theory,goldie2012non} and continues to suggests that direct feedline couplings at ``high" power might limit the performance of superconducting sensors.

\section{Conclusion}

We have constructed and tested an initial Al based Quantum Parity Detector, having embarked on an R\&D program to develop a sub-eV threshold phonon-mediated particle detector. These devices are fabricated using a relatively novel multi-step ion-mill process, and we have outlined both successes and deficiencies of this approach. We have determined that these QPDs are indeed capable of registering quasiparticle tunneling in an easily observable manner, and have further confirmed that their background quasiparticle density is both low, and in line with expectation.

\bibliographystyle{IEEEtran}
\bibliography{main.bib}

\end{document}